\documentclass[12pt,onecolumn,amssymb,floatfix,longbibliography]{revtex4-2}

\usepackage{amsmath}
\usepackage{color}
\usepackage{amsfonts}
\usepackage{amssymb}
\usepackage{graphicx}
\usepackage{geometry}
\usepackage{hyperref}
\usepackage{comment}
\usepackage{tabularx}
\usepackage{bm}
\usepackage{euscript}
\usepackage{graphicx}
\usepackage{color}
\usepackage{amsfonts}
\usepackage{exscale}
\usepackage{amsbsy}
\usepackage{textcomp}
\usepackage{comment}
\usepackage{hyperref}
\usepackage{slashed}
\usepackage{mathtools}
\usepackage{tabularx}
\usepackage{bm}
\usepackage{euscript}
\usepackage{graphicx}

\usepackage{color}
\usepackage{amsfonts}
\usepackage{exscale}
\usepackage{amsbsy}
\usepackage{subfig}
\usepackage{textcomp}
\usepackage{comment}
\usepackage{hyperref}
\usepackage[dvipsnames]{xcolor}
\usepackage{natbib}
\usepackage{enumitem}
\usepackage{ulem}

\newcommand{\nn}{\nonumber}

\numberwithin{equation}{section}

\begin{document}

\title{Interaction-enhanced transport crossover temperature in a Luttinger liquid}
\bigskip
\author{Yen-Wen Lu}
\author{Michael Mulligan}
\affiliation{Department of Physics and Astronomy, University of California, Riverside, California 92521, USA}
\date{\today}

 \bigskip
 \bigskip
 \bigskip
 \bigskip
 \bigskip
 \bigskip

\begin{abstract}
Strange metals are highly entangled gapless states of matter that exhibit anomalous transport, such as linear in temperature resistivity, over more than a decade of temperature.
Why a single power law should be so robust is an open question.
We propose a scenario in which interactions enhance the domain of certain scattering regimes, effectively suppressing other ``would-be regimes."
We test this proposal in a one-dimensional Luttinger liquid coupled to a one-dimensional acoustic phonon.
We use the memory matrix formalism to calculate the dc electrical and thermal conductivities at low and high temperatures, relative to the Debye cutoff on phonon frequencies, in both the ``clean" (umklapp scattering) and ``dirty" (disorder scattering) limits.
We find the crossover temperature separating the low and high temperature regimes to be interaction-dependent, with repulsive interactions increasing it, sometimes by more than an order of magnitude, depending on the scattering type.
This provides a concrete illustration for how interactions can extend a single transport regime over a wider temperature range.
\end{abstract}

\maketitle

\bigskip

\newpage

\thispagestyle{empty}


\newpage

\setcounter{page}{1}

\section{Introduction}

An infamous property of strongly correlated electron systems---from cuprate and heavy-fermion superconductors to twisted bilayer graphene---is robust $T$-linear electrical resistivity over unusually wide temperature ranges \cite{Carlson2008, RevModPhys.92.031001, Cao2020, RevModPhys.94.041002}. 
States that exhibit this behavior are called strange metals.
There are two features to distinguish here about strange metals: the first is the $T$-linear exponent; the second is the multiple orders of magnitude in temperature over which this exponent persists without apparent change. 
It is also noteworthy that this behavior occurs in a variety of disparate materials, seemingly ruling out an explanation that invokes fine-tuning.
Linear in $T$ resistivity can occur at low temperatures in non-Fermi liquids \cite{Varma1989, RevModPhys.94.035004, 2024PhRvL.132w6501B}---although there can be more conventional instances, e.g., in doped semiconductors \cite{PhysRevB.106.155427} or dilute metals \cite{Hwang2019} over limited temperature ranges---and may reflect a ``universal" Planckian scattering rate $\tau^{-1} \propto k_B T/\hbar$ \cite{sachdev2011book, Zaanen2019, RevModPhys.94.041002}.
This Planckian timescale represents a possible bound on dissipation in quantum systems.

In this paper, we are interested in the second feature of a strange metal: the persistence of a fixed exponent across a large temperature range.
This feature is particularly puzzling because different microscopic dissipation mechanisms should govern transport at different temperatures scales. 
Electron-phonon scattering provides the canonical example for this: In conventional 3d metals, it produces $T^5$ electrical resistivity at low temperatures and $T$-linear electrical resistivity at sufficiently high temperatures  \cite{Zimanbook}.
If electron-phonon scattering is also present in strongly correlated electron systems exhibiting robust $T$-linear resistivity, why does the resistivity slope in the high-temperature regime coincide with that at lower temperatures where quantum effects should prevail?

We propose that electron-electron interactions can modify the crossover temperature between low- and high-temperature transport regimes. 
Specifically, strong electron correlations may enhance the crossover scale, effectively extending the low-temperature regime to higher temperatures.
Within a suitable model, this mechanism could explain the persistence of non-Fermi liquid transport at elevated temperatures without invoking coincidental matching of scattering rates.

To test this proposal, we study transport in a one-dimensional Luttinger liquid coupled to a one-dimensional acoustic phonon, a system previously studied in \cite{Shimshoni2003,Seelig2005,Chudzinsk2020}. 
We use bosonization \cite{Giamarchibook} to treat certain electron-electron interactions exactly and the memory matrix formalism \cite{Giamarchibook, hartnoll2018holographic} to systematically compute transport coefficients across all temperatures.
We calculate both electrical $\sigma$ and thermal $\kappa$ conductivities in ``clean" (umklapp-dominated) and ``dirty" (disorder-dominated) limits, explicitly incorporating the Debye cutoff on phonon frequencies.
We assume a weak electron-phonon coupling and an arbitrarily strong, repulsive electron-electron interaction.

Our central observation is that the crossover temperature $T_0(K)$ separating the low- and high-temperature transport regimes is interaction-dependent.
Here, $K \leq 1$ parameterizes the effective interaction strength: electron-electron interactions vanish when $K=1$, while $K < 1$ (and decreasing) represents increasingly strong repulsive interactions.
Our observation applies to both electrical and thermal transport in the ``clean" and ``dirty" scattering limits.
The specific crossover temperature and transport exponents depend on both $K$ and the scattering regime.
For strong repulsive interactions ($K \ll 1$), we find that $T_0(K)$ is generally enhanced by more than an order of magnitude relative to the non-interacting case.

The remainder of the paper is organized as follows.
In \S \ref{modelsection} we introduce our conventions for describing a Luttinger liquid coupled to phonons.
In \S \ref{transportsection} we calculate the electrical and thermal conductivities of our system at low and high temperatures: we first introduce the memory matrix formalism; then we calculate these conductivities in the clean and dirty  limits. 
In \S \ref{crossoversection} we extract the crossover temperature $T_0(K)$ between low and high temperature regimes and observe how electron-electron interactions can enhance this temperature.
In \S \ref{conclusionsection} we conclude.
We provide the details of the calculations presented in the main text in a series of appendices. 

\section{Model}
\label{modelsection}

\subsection{Decoupled Fixed Point}

We consider a single channel of spinless electrons in one spatial dimension, with low-energy Hamiltonian:
\begin{align}
H_{LL} = H_{\rm lin}+H_{\rm int}.
\end{align}
The non-interacting Hamiltonian is
\begin{align}
H_{\rm lin}=-iv_{F}\int_{x}\big(\psi_{R}^{\dag}\partial_{x}\psi_{R}-\psi_{L}^{\dag}\partial_{x}\psi_{L}\big).
\end{align}
Here, $\psi_{R/L}^{\dag}$ creates a right/left-moving fermion with Fermi velocity $v_{F}$ about the Fermi point $\pm k_F$; we abbreviate $\int_{x}\equiv\int dx$. 
The interacting Hamiltonian consists of two terms,
\begin{align}
H_{\rm int} &= V_{f} + V_{b},
\end{align}
where the forward-scattering $V_{f}$ and backward-scattering $V_{b}$ terms,
\begin{align}
V_{f}=\pi g_{4}\int_{x}\big(\rho_{R}^{2}(x)+\rho_{L}^{2}(x)\big),\quad V_{b}=2\pi g_{2}\int_{x}\rho_{R}(x)\rho_{L}(x),
\end{align}
and $\rho_{R/L}(x)$ is the number density operator for right/left-movers. 
Repulsive interactions have $g_{2} > 0$.

The electron-electron scattering terms can be treated exactly using bosonization \cite{Giamarchibook}:
\begin{align}
H_{LL} & = v_{e} \int {dx \over 2 \pi} \left(K(\pi\Pi)^{2}+{1\over K}(\partial_{x}\phi)^{2}\right),
\end{align}
where the canonical momentum $\Pi=(\partial_{x}\theta)/\pi$,
\begin{align}
v_{e}&={1 \over 4}\sqrt{(v_{F}+g_{4})^{2}-g_{2}^{2}},\quad K=\left[{v_{F}+g_{4}-g_{2}\over v_{F}+g_{4}+g_{2}}\right]^{1\over 2}.
\end{align}
Forward scattering $g_4$ can be absorbed into a redefinition of the Fermi velocity $v_F$.
The Luttinger parameter $K < 1$ for repulsive interactions ($g_{2} > 0$).
Microscopically, small values of $K$ are realized in systems with long-ranged Coulomb interactions \cite{Schulz1993,Glazman1993,Giamarchibook}.
Notice that the Luttinger velocity $v_e \rightarrow 0$ as $K \rightarrow 0$.
We therefore require $K > 0$.
The right/left-moving fermion operators are related to the Luttinger bosons ($\phi$ and $\theta$) as
\begin{align}
\psi_{R/L}(x)={1\over\sqrt{2\pi a}}e^{i(\pm\phi-\theta)},
\end{align}
where $\phi=(\varphi_{L}+\varphi_{R})/2$ and $\theta=(\varphi_{L}-\varphi_{R})/2$, with right/left-moving bosons $\varphi_{R/L}$; $a > 0$ is a short-distance cutoff.

One-dimensional acoustic phonons have the Hamiltonian,
\begin{align}
\label{phononhamiltonian}
H_{ph} &= {1 \over 2 \pi} \int_x \left[(\pi P)^{2} + v_{p}^{2}(\partial_{x} q)^{2}\right].
\end{align}
Here $P$ is the canonical momentum for the phonons with ``coordinate" (or displacement field) $q$ and $v_p$ is the velocity of the phonon modes.
We assume $v_p \ll v_e$.

The decoupled fixed point Hamiltonian for the Luttinger boson + acoustic phonon system is
\begin{align}
\label{fixedpointhamiltonian}
H_{0} = H_{LL} + H_{ph}.
\end{align}
This fixed point Hamiltonian has an infinite number of conserved quantities.
We are interested in the ``slowly-decaying" symmetries associated with translation invariance, charge (or fermion number) conservation, and energy conservation that are preserved when the Luttinger boson and acoustic phonon are coupled together.
These symmetries have the following conserved currents:
 the momentum operator, 
\begin{align}
P_{D}&=\int_{x} \Pi(\partial_{x}\phi)+\int_{x} P(\partial_{x}q),
\end{align}
or, equivalently, the ``momentum current" $J_D = - v_e^2 P_D$;
the electrical current, 
\begin{align}
J_{e}=ev_{e}K\int_{x}\Pi;
\end{align}
and the heat current,
\begin{align}
J_{T}=-v_{e}^{2}\int_{x} \Pi(\partial_{x}\phi)-v_{p}^{2}\int_{x} P(\partial_{x}q).
\end{align}

\subsection{Current Relaxation}

The currents $\{ J_D, J_e, J_T \}$ commute with $H_0$ in Eq.\eqref{fixedpointhamiltonian}.
Consequently, the decoupled fixed point has infinite conductivity \cite{PhysRevB.55.11029}.
Finite conductivity requires current relaxation and momentum dissipation, i.e., we need to supplement $H_0$ with terms that do not commute with $\{ J_D, J_e, J_T \}$.

We will study two different types of current relaxation mechanisms that couple the Luttinger boson and acoustic phonon:
umklapp scattering, 
\begin{align}
\label{umklappscattering}
H^{U} = \sum_{m \geq 1} H_{m}^{U}  =  -\sum_{m} \lambda_{m}^{U} \int_{x}\big[ {1 \over a^{m}} e^{i \overline{k}_m  x} e^{im \phi} \partial_{x} q  + {\rm h.c.} \big];
\end{align}
and disorder scattering,
\begin{align}
\label{disorderscattering}
H^{{\rm dis}} = \sum_{m \geq 1} H_{m}^{{\rm dis}} = \sum_{m} \lambda_{m}^{{\rm dis}} \int_{x}\big[ {1 \over a^{m}} \xi_{m}(x) e^{im\phi} \partial_{x} q  + {\rm h.c.} \big].
\end{align}
These two types of scattering correspond to ``clean" and ``dirty" current relaxations.
Intuitively, the vortex operator $e^{im \phi} \sim(\psi_{R} \psi_{L}^\dagger)^{m}$, where $m \in\mathbb{N}^{+}$.
The momentum mismatch $\overline{k}_m$ in \eqref{umklappscattering} equals
\begin{align}
\overline{k}_m = m k_{F} - p_m G\in[0,2\pi),
\end{align}
where $G$ is the reciprocal lattice vector (assuming an underlying lattice).
For commensurate fillings, there is an integer $p$ such that $\overline{k}_m = 0$.
$\xi_m(x)$ is a quenched random variable with Gaussian statistics:
\begin{align}
\overline{\xi_{m}(x)}=0,\qquad\overline{\xi_{m}(x)\xi_{m'}^{*}(x')}=D_m \delta_{m m'}\delta(x-x'),
\end{align}
where the overline represents the disorder average. 

$H_m^{U}$ and $H_m^{\rm dis}$ are scattering processes between $m$ left-moving and $m$ right-moving fermions.
The leading processes occur at $m = 1$.
$e^{im \phi}$ has scaling dimension $\Delta_m (K) = m^2 K$.
For commensurate fillings, $\lambda_m^U$ has (leading) beta function $\beta^U_{m} = \big( 2 - (\Delta_m + 1) \big) \lambda_m^U = \big( 1 - \Delta_m \big) \lambda_m^U$.
(The $``1"$ in $(\Delta_m + 1)$ arises the from the spatial derivative on the dimensionless phonon field $q$.)
Thus, $\lambda_m^U$ is relevant for $\Delta_m < 1$; this occurs for any repulsive interaction ($K < 1$) at $m=1$.
We use the convention that a positive beta function corresponds to a relevant operator and a negative beta function corresponds to an irrelevant operator.
For incommensurate fillings, the exponential term $e^{i \overline{k}_m x}$ renders $\lambda_{m}^{U}$ effectively zero at long distances.
Upon performing the usual (replicated) disorder average \cite{Giamarchibook}, the variance $D_m$ has (leading) beta function $\beta^{\rm dis}_{m} = (3 - 2 (\Delta_m + 1)) D_m = \big( 1 - 2 \Delta_m \big) D_m$.
Thus, the disorder scattering is irrelevant for $\Delta_m > 1/2$ and relevant for $\Delta_m < 1/2$.
In our calculation, we {\it assume} both types of scattering processes are weak. 
Given the above analysis, this occurs with umklapp scattering at incommensurate fillings or with disorder scattering for $1/2 < K \leq 1$.

\section{Transport at low and high temperatures}
\label{transportsection}

\subsection{Memory Matrix}

The memory matrix $\hat{{\cal M}}(\omega)$ is a convenient way to encode the transport properties of any system (see \cite{hartnoll2018holographic} for a modern review of the formalism in general and \cite{Giamarchibook} for a discussion in the context of the Luttinger liquid).
Its definition and relation to the electrical and thermal conductivities proceed as follows:
\begin{align}
\hat{{\cal M}}(\omega) & = \sum_{m}\big( \hat{\cal M}_{m}^{U}(\omega) + \hat{\cal M}_{m}^{{\rm dis}}(\omega) \big),
\end{align}
where the contribution of each scattering process is
\begin{align}\label{memorydef1}
(\hat{\cal M}^{U}_{m})^{{\cal Q} {\cal Q}'} & = {1\over L} {\langle F^{U}_{m,{\cal Q}} ; F^{U}_{m,{\cal Q}'} \rangle_{\omega} - \langle F^{U}_{m,{\cal Q}} ; F^{U}_{m,{\cal Q}'} \rangle_{\omega=0} \over i\omega},\\
(\hat{\cal M}^{{\rm dis}}_{m})^{{\cal Q} {\cal Q}'} & = {1\over L} {\langle F^{{\rm dis}}_{m,{\cal Q}} ; F^{{\rm dis}}_{m , {\cal Q}'} \rangle_{\omega} - \langle F^{{\rm dis}}_{m , {\cal Q}} ; F^{{\rm dis}}_{m , {\cal Q}'} \rangle_{\omega=0} \over i\omega}.\label{memorydef2}
\end{align}
Here, $F_{\alpha,{\cal Q}}^{U} = i[H_{\alpha}^{U} , {\cal Q}]$, $F_{\alpha,{\cal Q}}^{{\rm dis}} = {i \over \sqrt{D_{m}}}[H_{\alpha}^{{\rm dis}},{\cal Q}]$, and the conserved current $\mathcal{Q} \in \{ J_D, J_e, J_T \}$. 
$\langle F^{U}_{m,{\cal Q}} ; F^{U}_{m,{\cal Q}'} \rangle_{\omega}$ and $\langle F^{{\rm dis}}_{m,{\cal Q}} ; F^{{\rm dis}}_{m,{\cal Q}'} \rangle_{\omega}$ are finite-temperature retarded Green's functions, evaluated at real frequency $\omega$. 
Notice that $\hat {\cal M}_m(\omega)$ has a zero eigenvalue for any current that commutes with $H^U_m$ (or $H^{\rm dis}_m$).
The static susceptibility matrix $\hat{\chi}$ is the retarded finite-temperature Green's function:
\begin{align}
\label{staticsusceptibilities}
\hat{\chi}_{{\cal Q} {\cal Q}'}=({\cal Q}|{\cal Q}')\equiv{1\over L}G_{{\cal Q} {\cal Q}'}(\omega=0).
\end{align}
The static susceptibility defines the overlap of two currents. 
Finite overlap between two currents implies that their transport is correlated.

The conductivity matrix $\hat{\sigma}(\omega)$ is defined in terms of the memory matrix $\hat{{\cal M}}$ and static susceptibility $\hat{\chi}$:
\begin{align}\label{transport}
\hat{\sigma}(\omega)=\hat{\chi}(\hat{{\cal M}}(\omega)-i\omega\hat{\chi})^{-1}\hat{\chi}.
\end{align}
Intuitively, the memory matrix formalism first computes the resistivity and then inverts to get the conductivity.
This order of computation is especially useful for our system, with infinite conductivity in the decoupled limit.
The electrical conductivity $\sigma(\omega)$ is the submatrix $\hat{\sigma}(\omega)_{J_{e}J_{e}}$ and the thermal conductivity $\kappa(\omega)=\hat{\sigma}(\omega)_{J_T J_T}/T$. 
The thermoelectric conductivity $\widetilde{\alpha}$ is always $0$ in our work, since there is no overlap between $J_{e}$ and $J_{T}$ or $J_{D}$, i.e., $(J_{e}|J_{T})=(J_{e}|J_{D})=0$. 
We get the dc electrical $\sigma$ and thermal $\kappa$ conductivities by taking $\omega \rightarrow 0$.

\subsection{Umklapp scattering}
\label{umklappsection}

In this section, we calculate the conductivity matrix $\hat \sigma$ \eqref{transport} due to umklapp scattering \eqref{umklappscattering}.
Similar low-temperature conductivity calculations have been performed in \cite{Shimshoni2003,Seelig2005,Chudzinsk2020}.

We begin with the static susceptibilities \eqref{staticsusceptibilities} in the limit of ${v_{p}/v_{e}}\ll 1$:
\begin{align}
\hat{\chi}_{J_{e}J_{e}}&={e^{2}v_{e}K\over \pi},\\
\hat{\chi}_{J_{T}J_{T}} &={v_{p}^{3}\over v_{e}^{3}}\hat{\chi}_{J_{D} J_{D}}={v_{p}\over v_{e}}\hat{\chi}_{J_{D} J_{T}}={v_{p}\over v_{e}}\hat{\chi}_{J_{T} J_{D}}={\pi v_{e}T^{2}\over 3},\\
\hat{\chi}_{J_{e} J_{T}} &= \hat{\chi}_{J_{T} J_{e}} = \hat{\chi}_{J_{e} J_{D}} = \hat{\chi}_{J_{D} J_{e}} = 0.
\end{align}
The calculation of these Green's functions is standard, with all details relegated to Appendix \ref{appendixsusceptibility}.
Note that the static susceptibilities are independent of the particular relaxation mechanism and so these susceptibilities will also be used in the next section when we consider disorder scattering.

We next turn to the memory matrix $(\hat{\cal M}^{U}_{m})^{{\cal Q} {\cal Q}'}$, where $\mathcal{Q} \in \{ J_D, J_e, J_T \}$.
According to \eqref{memorydef1} and using results in Appendix \ref{commute}, we need the retarded Green's functions:
\begin{align}\label{memory1}
{1\over L}\langle F_{m,J_{e}}^{U};F_{m,J_{e}}^{U}\rangle_{\omega}& \propto {1\over L}\int\limits_{t,x_{+},x_{-}}e^{i(\omega t-\overline{k}_{m}x_{-})}g_{e}(x_{-},t)g_{p}(x_{-},t)+(\overline{k}_{m}\rightarrow -\overline{k}_{m}),\\
{1\over L}\langle F_{m, {\cal P}}^{U};F_{m,{\cal P}'}^{U}\rangle_{\omega}& \propto {1\over L}\int\limits_{t,x_{+},x_{-}}e^{i(\omega t-\overline{k}_{m}x_{-})}g_{e}(x_{-},t)\left(\partial_{x_{-}}^{2}g_{p}(x_{-},t)\right)+(\overline{k}_{m}\rightarrow -\overline{k}_{m}),\label{memory2}
\end{align}
where ${\cal P}, {\cal P}' \in\{J_{T},J_{D}\}$.
We have simplified the presentation of the above expressions by leaving out constant proportionality factors. 
The real-space fermion two-point function $g_{e} (x,t)$ is
\begin{align}
g_{e}(x_{-},t) &\equiv \left\langle {e^{\pm im \phi(x,t)}\over a^{m }} {e^{\mp im \phi(y,0)}\over a^{m}} \right\rangle =e^{C_{e}(x_{-},t)},\label{electroncorr}
\end{align}
where
\begin{align}
C_{e}(x,t)&\equiv \Delta_m(K)\ln\left[{\pi aT/v_{e}\over\sinh(\pi T(x-v_{e}t+ia)/v_{e})}\right] + \Delta_m(K)\ln\left[{\pi aT/v_{e}\over\sinh(\pi T(x+v_{e}t-ia)/v_{e})}\right],
\end{align}
using the abbreviation $x_{\pm}=x\pm y$.
Recall that $\Delta_m(K) = m^2 K$. 
The phonon two-point function $g_p(x,t)$ is 
\begin{align}
g_{p}(x_{-},t)&\equiv\left\langle \partial_{x} q(x,t) \partial_{y} q(y,0) \right\rangle = \int_k {k^2 \over 2 \omega_k} e^{ikx}\left\{N(\omega_{k})e^{i\omega_{k}t}+[N(\omega_{k})+1]e^{-i\omega_{k}t}\right\}, \label{phononcorr}\cr
\end{align}
where $\int_k \equiv \int {dk \over 2 \pi}$ and $N(\omega_{k})=(e^{\omega_{k}/T}-1)^{-1}$ is phonon occupation, with $\omega_k = v_p |k|$.
The integral over phonon momentum is cut off at $|k| = k_D = \Theta_D/v_p$, with $\Theta_D$ the Debye temperature. 

The integrals over $x_+$ on the right-hand sides of \eqref{memory1} and \eqref{memory2} are trivially done and give a factor of $L$, which cancels the $L^{-1}$ coefficients in the two expressions.
What remains are essentially Fourier transforms of products of correlation functions. 
To simplify subsequent expressions, we drop ``$-$" subscript: $x_- \rightarrow x$. 

Substituting the right-hand sides of \eqref{memory1} and \eqref{memory2} (after performing the $x_+$ integrals) into the expressions for the memory matrix $({\cal \hat{M}}^{U}_{m})^{{\cal Q} {\cal Q}'}$ \eqref{memorydef1}, we obtain, in the dc limit:
\begin{align}
\lim_{\omega \rightarrow 0} ({\cal \hat{M}}^{U}_{m})^{J_{e}J_{e}}&= \lim_{\omega \rightarrow 0} A_{m,J_{e}}^{U}A_{m,J_{e}}^{U}{G_{1}(\overline{k}_m ,\omega)-G_{1}(\overline{k}_m,\omega=0)\over i\omega}+(\overline{k}_m \rightarrow - \overline{k}_m)\label{dc1}\cr
&= \left.A_{m,J_{e}}^{U}A_{m,J_{e}}^{U}\big(-i\partial_{\omega}G_{1}(\overline{k}_m,\omega)\big)\right|_{\omega=0}+(\overline{k}_m \rightarrow - \overline{k}_m),\\
\lim_{\omega \rightarrow 0} ({\cal \hat{M}}^{U}_{m})^{ {\cal P} {\cal P}'}&=\lim_{\omega \rightarrow 0} A_{m,{\cal P}}^{U}A_{m,{\cal P}'}^{U}{\widetilde{G}_{1}(\overline{k}_m ,\omega)-\widetilde{G}_{1}(\overline{k}_m,\omega=0)\over i\omega}+(\overline{k}_m \rightarrow - \overline{k}_m)\label{dc2}\cr
& = \left.A_{m,{\cal P}}^{U}A_{m,{\cal P}'}^{U}\big(-i\partial_{\omega}\widetilde{G}_{1}(\overline{k}_m,\omega)\big)\right|_{\omega=0}+(\overline{k}_m \rightarrow - \overline{k}_m).
\end{align}
The constant coefficients $A_{m,{\cal Q}}^{i}$ are given in Appendix \ref{commute} and the $G$-functions are:
\begin{align}
G_{1}(\bar{k},\omega)&=\int\limits_{x,t}e^{i(\omega t-\bar{k}x)}g_{e}(x,t)g_{p}(x,t)\label{electrical}\\
&=\int_k {k^{2}\over 2 \omega_{k}}\left\{N(\omega_{k})G_{e}(\bar{k}-k,\omega+\omega_{k})+[N(\omega_{k})+1]G_{e}(\bar{k}-k,\omega-\omega_{k})\right\},\cr
\widetilde{G}_{1}(\bar{k},\omega)&=-\int\limits_{x,t}e^{i(\omega t-\bar{k}x)}g_{e}(x,t)\big(\partial_{x}^{2}g_{p}(x,t)\big)\label{thermal}\\
&=\int_k {k^{4}\over 2 \omega_{k}}\left\{N(\omega_{k})G_{e}(\bar{k}-k,\omega+\omega_{k})+[N(\omega_{k})+1]G_{e}(\bar{k}-k,\omega-\omega_{k})\right\}. \nn
\end{align}
The second lines in the above two expressions are obtained by directly substituting in the phonon two-point function \eqref{phononcorr} and introducing the Fourier transform of the fermion two-point function, $G_{e}(k,\omega)$.
It turns out that $G_{e}(k,\omega)$ can be computed exactly (see, e.g., (C.68) in \cite{Giamarchibook}):
\begin{align}
G_{e}(k,\omega)&=\int\limits_{x,t}e^{i(\omega t-kx)}g_{e}(x,t)\label{correlation} \\
& =-{a^{2\Delta_m}\sin(\pi \Delta_m)\over v_{e}}\big({2\pi T\over v_{e}}\big)^{2\Delta_m-2} \cr
& \times B({\Delta_m\over 2}-i{(\omega-kv_{e})\over 4\pi T},1-\Delta_m)B({\Delta_m\over 2}-i{(\omega+kv_{e})\over 4\pi T},1-\Delta_m),
\end{align}
where $B(\alpha, \beta) = \Gamma(\alpha) \Gamma(\beta)/\Gamma(\alpha + \beta)$ is the beta function. 

In the limit $v_{p}\ll v_{e}$, $G_{e}(\bar{k}-k,\omega\pm\omega_{k})\approx G_{e}(\bar{k}-k,\omega)$.
Thus, \eqref{electrical}) and \eqref{thermal} simplify to
\begin{align}
G_{1}(\bar{k},\omega)& = \int_k {k^{2}\over 2 \omega_{k}} [2N(\omega_{k})+1]G_{e}(\bar{k}-k,\omega),\\
\widetilde{G}_{1}(\bar{k},\omega)& = \int_k{k^{4}\over 2 \omega_{k}} [2N(\omega_{k})+1]G_{e}(\bar{k}-k,\omega).
\end{align}
Since $\omega_{k}$ and $N(\omega_{k})$ are symmetric under $k\rightarrow -k$,
\begin{align}
\label{G1simple}
G_{1}(\bar{k},\omega)&= \int_0^{k_D} {dk \over 2\pi} {k^{2}\over 2 \omega_{k}} [2N(\omega_{k})+1]\left[G_{e}(\bar{k}-k,\omega)+G_{e}(\bar{k}+k,\omega)\right],\\
\label{G1tildesimple}
\widetilde{G}_{1}(\bar{k},\omega)&=\int_0^{k_D} {dk \over 2\pi} {k^{4}\over 2 \omega_{k}} [2N(\omega_{k})+1]\left[G_{e}(\bar{k}-k,\omega)+G_{e}(\bar{k}+k,\omega)\right],
\end{align}
where we have explicitly indicated the phonon momentum cutoff $k_D = \Theta_D/v_p$.

It remains to evaluate the dc limit of the memory matrix, \eqref{dc1} and \eqref{dc2}, after plugging in the $G$-functions, \eqref{G1simple} and \eqref{G1tildesimple}, and the $v_p \ll v_e$ limit of the (Fourier-transformed) fermion two-point function \eqref{correlation}.
Let us separately consider the electrical and thermal parts of the memory matrix.
At low temperatures $T \ll \Theta_D$, the electrical component of the dc memory matrix is found to be 
\begin{align}
\lim_{\omega\rightarrow 0}({\cal \hat{M}}^{U}_{m})_{<}^{J_{e}J_{e}}&= A_{m,J_{e}}^{U}A_{m,J_{e}}^{U}{a^{2\Delta_m}\over v_{e}^{2}v_{p}}\big({2\pi T\over v_{e}}\big)^{2\Delta_m-1}{2^{4-2\Delta_m}\over \pi}\Gamma^{2}(1-\Delta_m)\sin^{2}(\Delta_m\pi)\Gamma(2\Delta_m+1)e^{-{\overline{k}_{m}v_{e}/2T}}.
\end{align}
Notice the exponential suppression as $T \rightarrow 0$ for incommensurate fillings $\overline{k}_{m} \neq 0$, as found in \cite{Shimshoni2003}.
At high temperatures $T\gg\Theta_{D}$, we find 
\begin{align}
\label{electricalhigh}
\lim_{\omega\rightarrow 0}({\cal \hat{M}}_{m}^{U})_{>}^{J_{e}J_{e}}&= A_{m,J_{e}}^{U}A_{m,J_{e}}^{U}{a^{2\Delta_m}\over v_{e}}\big({2\pi T\over v_{e}}\big)^{2\Delta_m-2}{2\over \pi}{\Theta_{D}\over v_{p}^{3}}\left|B({\Delta_m\over 2},1-\Delta_m)\right|^{2}\cos^{2}({\Delta_m\pi\over 2}).
\end{align}
Using these memory matrices and the static susceptibilities \eqref{staticsusceptibilities} (given at the start of this section), we finally obtain the dc electrical conductivity \eqref{transport} due to umklapp scattering: \begin{align}
\sigma\propto
\begin{cases}
T^{1-2\Delta_m}e^{|\overline{k}_m v_{e}|/2T},&T\ll\Theta_{D},\cr
T^{2-2\Delta_m},&T\gg\Theta_{D}.
\end{cases}
\end{align}

We repeat the above steps for the thermal component of the memory matrix.
At low temperatures, the thermal components of the memory matrix are
\begin{align}
\lim_{\omega\rightarrow 0}({\cal \hat{M}}_{m}^{U})_{<}^{{\cal P} {\cal P}'}=A_{m,{\cal P}}^{U}A_{m,{\cal P}'}^{U}{a^{2\Delta_m}\over v_{e}^{2}v_{p}}\big({2\pi T\over v_{e}}\big)^{2\Delta_m+1}{2^{4-2\Delta_m}\over \pi^{3}}\Gamma^{2}(1-\Delta_m)\sin^{2}(\Delta_m\pi)\Gamma(2\Delta_m+3)e^{-{\overline{k}_{m}v_{e}/2T}}.
\end{align}
At high temperatures, we find
\begin{align}
\lim_{\omega\rightarrow 0}({\cal \hat{M}}_{m}^{U})_{>}^{{\cal P} {\cal P}'}&=A_{m,{\cal P}}^{U}A_{m,{\cal P}'}^{U}{a^{2\Delta_m}\over v_{e}}\big({2\pi T\over v_{e}}\big)^{2\Delta_m-2}{2\over 3\pi}{\Theta_{D}^{3}\over v_{p}^{5}}\left|B({\Delta_m\over 2},1-\Delta_m)\right|^{2}\cos^{2}({\Delta_m\pi\over 2}).
\end{align}
Assembling these memory matrices with the static susceptibilities, we obtain the dc thermal conductivity due to umklapp scattering:
\begin{align}
\kappa\propto
\begin{cases}
T^{2-2\Delta_m}e^{|\overline{k}_m v_{e}|/2T},&T\ll\Theta_{D},\cr
T^{5-2\Delta_m},&T\gg\Theta_{D}.
\end{cases}
\end{align}
The Lorentz ratio is
\begin{align}
{\cal L}={\kappa\over T\sigma}\propto
\begin{cases}
{\rm constant},&T\ll\Theta_{D}, \cr
T^{2},&T\gg\Theta_{D}.
\end{cases}
\end{align}

\subsection{Disorder scattering}
\label{disordersection}

In this section, we present the conductivity matrix $\hat \sigma$ \eqref{transport} due to disorder scattering \eqref{disorderscattering}.
The logic of the calculation is the same as the previous section so we relegate all details of the calculation to Appendix \ref{appendixdisordermemory}. 

The dc electrical memory matrix is
\begin{align}
\lim_{\omega\rightarrow 0}( {\cal \hat{M}}_{m}^{\rm dis} )_{<}^{J_{e}J_{e}}&= A_{m, J_{e}}^{\rm dis} A_{m, J_{e}}^{\rm dis} {a^{2\Delta_m}\over v_{e}^{2}v_{p}}\big({2\pi T\over v_{e}}\big)^{2\Delta_m-2}\big({2T\over v_{p}}\big)^{2} e^{i(\Delta_m-1/2)\pi} {\pi^{2-2\Delta_m}\over \Gamma(2\Delta_m)}{\Gamma(2\Delta_m+1)\over 2^{2\Delta_m-1}},\\
\lim_{\omega\rightarrow 0}( {\cal \hat{M}}_{m}^{\rm dis} )_{>}^{J_{e}J_{e}}&= A_{m, J_{e}}^{\rm dis} A_{m, J_{e}}^{\rm dis} {a^{2\Delta_m}\over v_{e}^{2}v_{p}}\big({2\pi T\over v_{e}}\big)^{2\Delta_m-2}\big({2T\over v_{p}}\big)^{2}e^{i(\Delta_m-1/2)\pi}{|\Gamma(\Delta_m)|^{2}\over \Gamma(2\Delta_m)}\big({\Theta_{D}\over T}\big).
\end{align}
Thus, the dc electrical conductivity $\sigma$ is
\begin{align}
\sigma\propto
\begin{cases}
T^{-2\Delta_m},&T\ll\Theta_{D},\cr
T^{1-2\Delta_m},&T\gg\Theta_{D}.
\end{cases}
\end{align}

For ${\cal P}, {\cal P}'\in\{ J_{T},J_{D} \}$, the dc thermal memory matrix at low ($T \ll \Theta_D$) and high ($T \gg \Theta_D$) temperatures is found to be
\begin{align}
\lim_{\omega\rightarrow 0}( {\cal \hat{M}}_{m}^{\rm dis} )_{<}^{{\cal P} {\cal P}'} &= A_{m, {\cal P}}^{\rm dis} A_{m, {\cal P}'}^{\rm dis} {a^{2\Delta_m}\over v_{e}^{2}v_{p}}\big({2\pi T\over v_{e}}\big)^{2\Delta_m-2}\big({2T\over v_{p}}\big)^{4}e^{i(\Delta_m-1/2)\pi} {\pi^{2-2\Delta_m}\over \Gamma(2\Delta_m)}{\Gamma(2\Delta_m+2)\over 2^{2\Delta_m+2}},\\
\lim_{\omega\rightarrow 0}( {\cal \hat{M}}_{m}^{\rm dis} )_{>}^{{\cal P} {\cal P}'} &= A_{m, {\cal P}}^{\rm dis} A_{m, {\cal P}'}^{\rm dis} {a^{2\Delta_m}\over v_{e}^{2}v_{p}}\big({2\pi T\over v_{e}}\big)^{2\Delta_m-2}\big({2T\over v_{p}}\big)^{4}e^{i(\Delta_m-1/2)\pi}{|\Gamma(\Delta_m)|^{2}\over \Gamma(2\Delta_m)}{1\over 3}\big({\Theta_{D}\over T}\big)^{3}.
\end{align}
The resulting dc thermal conductivity $\kappa$ scales as
\begin{align}
\kappa\propto
\begin{cases}
T^{1-2\Delta_m},&T\ll\Theta_{D}, \cr
T^{4-2\Delta_m},&T\gg\Theta_{D}.
\end{cases}
\end{align}
The Lorentz ratio is
\begin{align}
{\cal L}={\kappa\over T\sigma}\propto
\begin{cases}
{\rm constant}, &T\ll\Theta_{D}\cr
T^{2}, & T\gg\Theta_{D}.
\end{cases}
\end{align}
The Lorentz ratio for disorder scattering is also non-constant, similar to what we found for the umklapp scattering.

\section{Crossover temperature}
\label{crossoversection}

\begin{figure}[t]
\includegraphics[height=3in,width=4in] {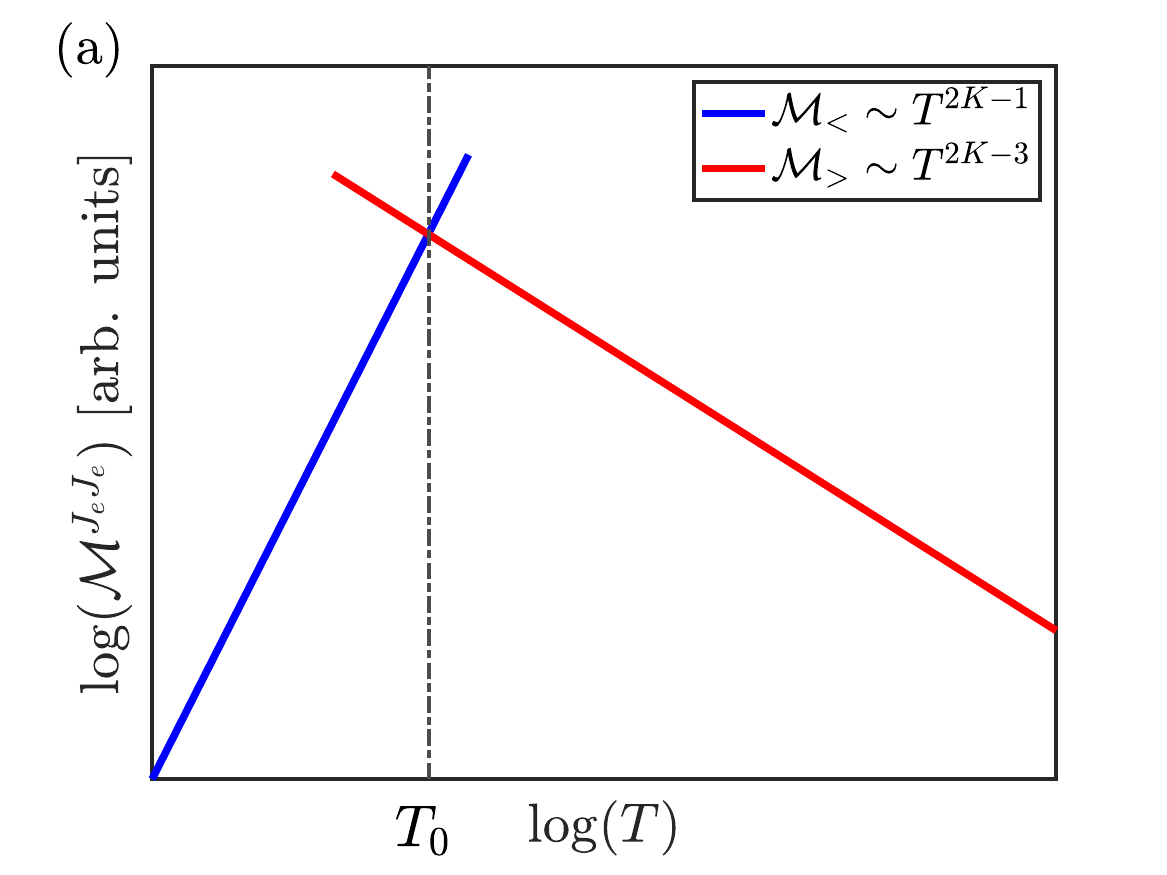}
\includegraphics[height=3in,width=4in] {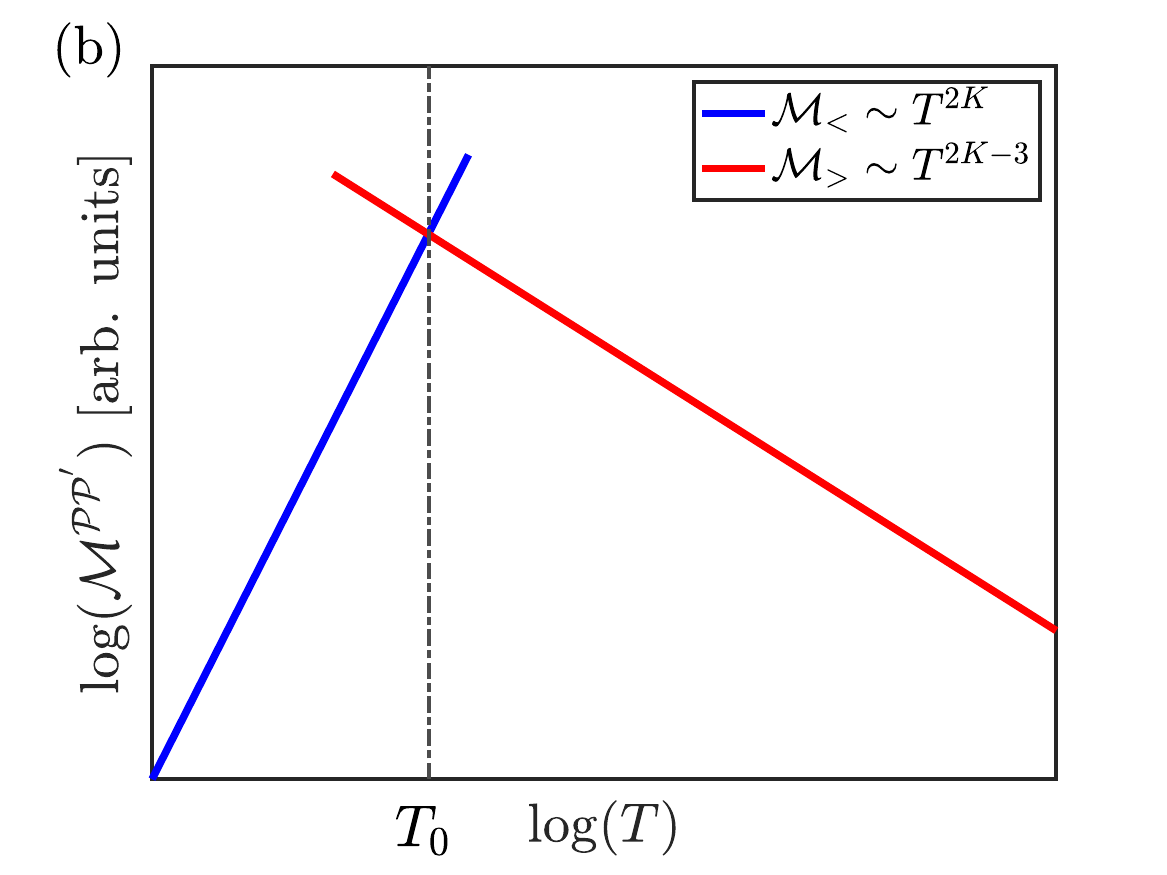}
\centering
\caption{Low- and high-temperature asymptotics of the memory matrix ${\cal M}$ with intervening crossover temperature $T_0$ in the (a) clean (unklapp-dominated) and (b) dirty (disorder-dominated) limits.
Blue (Red) lines are low (high)-temperature asymptotics of the memory matrices, and ${\cal P}, {\cal P'}\in\{J_{T},J_{D}\}$. }
\label{memorycross}
\end{figure}
In the previous section we presented the conductivity matrix for umklapp and disorder scattering processes \eqref{umklappscattering} and \eqref{disorderscattering}, for general $m$.
We now specialize to the leading scattering term, with $m=1$, corresponding to the phonon-mediated scattering between a single left- and right-moving fermion.
Thus, we replace $\Delta_m \rightarrow \Delta_{m=1} = K$.

The crossover temperature $T^{i, j}_0$ is where the low and high temperature transport asymptotics coincide 
(see Fig.~\ref{memorycross}):
\begin{align}
\label{crossovertemperaturedefinition}
\lim_{\omega\rightarrow 0}({\cal \hat{M}}_{m}^{i})_{<}^{{\cal Q} {\cal Q}'} = \lim_{\omega\rightarrow 0}({\cal \hat{M}}_{m}^{i})_{>}^{{\cal Q} {\cal Q}'}.
\end{align}
Here, the superscripts on $T_0^{i, j}$ indicate the scattering mechanism $i \in \{U, {\rm dis} \}$ and the type of transport $j \in \{e, {\rm th} \}$.
For example, $T_0^{U, e}$ is the crossover temperature for the electrical resistivity due to umklapp scattering.

Solving \eqref{crossovertemperaturedefinition} for $T^{i,j}_0$, we find
\begin{align}
\label{crossovertemperatureinteractions}
T^{U,j}_0 & = T^{U,j}_\ast\left[2^{2K-2}{\Gamma(2+\alpha)\over\Gamma(2K +\alpha)}\left({\Gamma({K \over 2})\over\Gamma({1\over 2})}\right)^{4}\right]^{1/\alpha},\\
\label{crossovertemperatureinteractionsthermal}
T^{{\rm dis},j}_0 & = T^{{\rm dis},j}_\ast\left[(2\pi)^{2K-2}{\Gamma(2+\alpha)\over\Gamma(2K +\alpha)}\Gamma^{2}(K)\right]^{1/\alpha},
\end{align}
where $\alpha = 1$ when $j = e$, $\alpha = 3$ when $j = {\rm th}$.
The constant coefficient $T^{i, j}_\ast$ is the value of the crossover temperature at the noninteracting point $K = 1$:
\begin{align}
T^{U,j}_\ast &=\left[{2^{-\alpha-4}\over\alpha\pi}{1\over\Gamma(2+\alpha)}\left({v_{e}\over v_{p}}\right)^{2\alpha+1}\right]^{1/\alpha}\Theta_{D},\\
T^{{\rm dis},j}_\ast &= \left[{1\over\alpha}{1\over\Gamma(2+\alpha)}\right]^{1/\alpha}\Theta_{D}.
\end{align}

\begin{figure}[t]
\includegraphics[height=3in,width=4in] {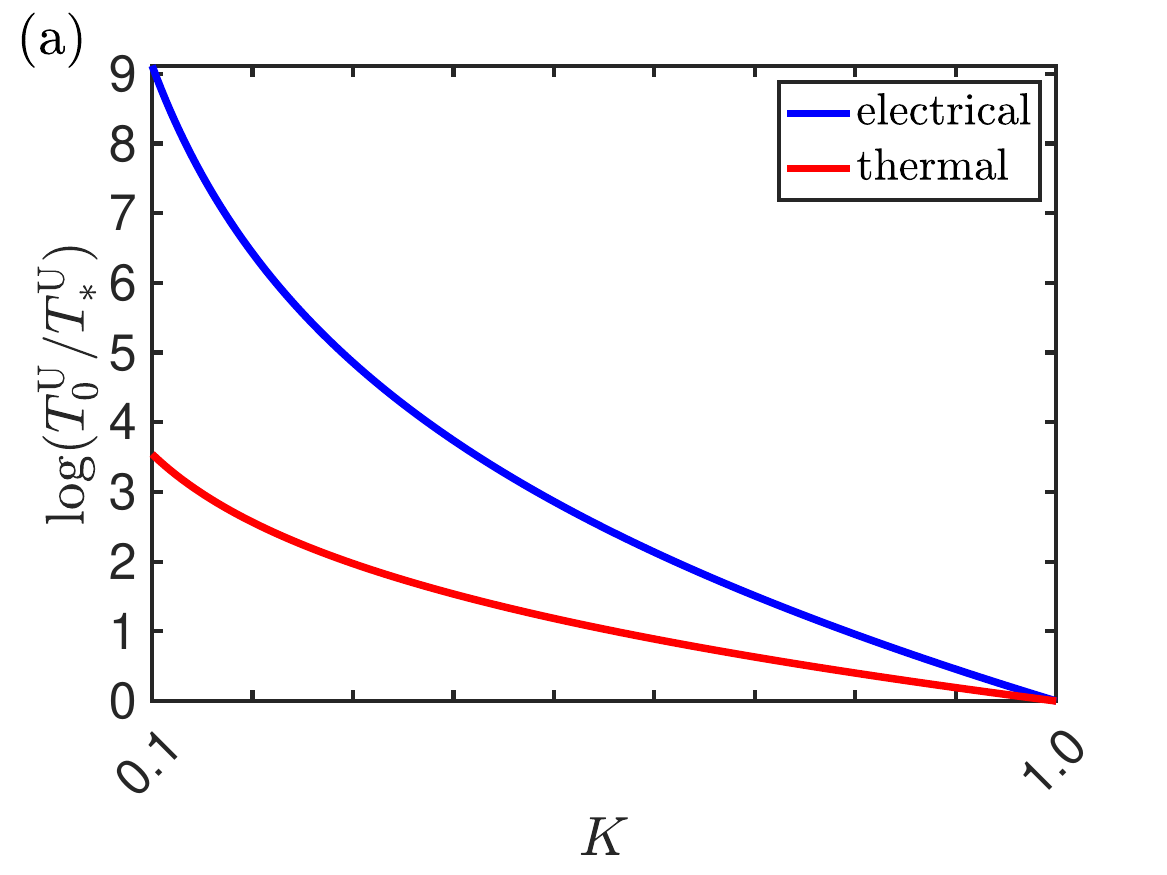}
\includegraphics[height=3in,width=4in] {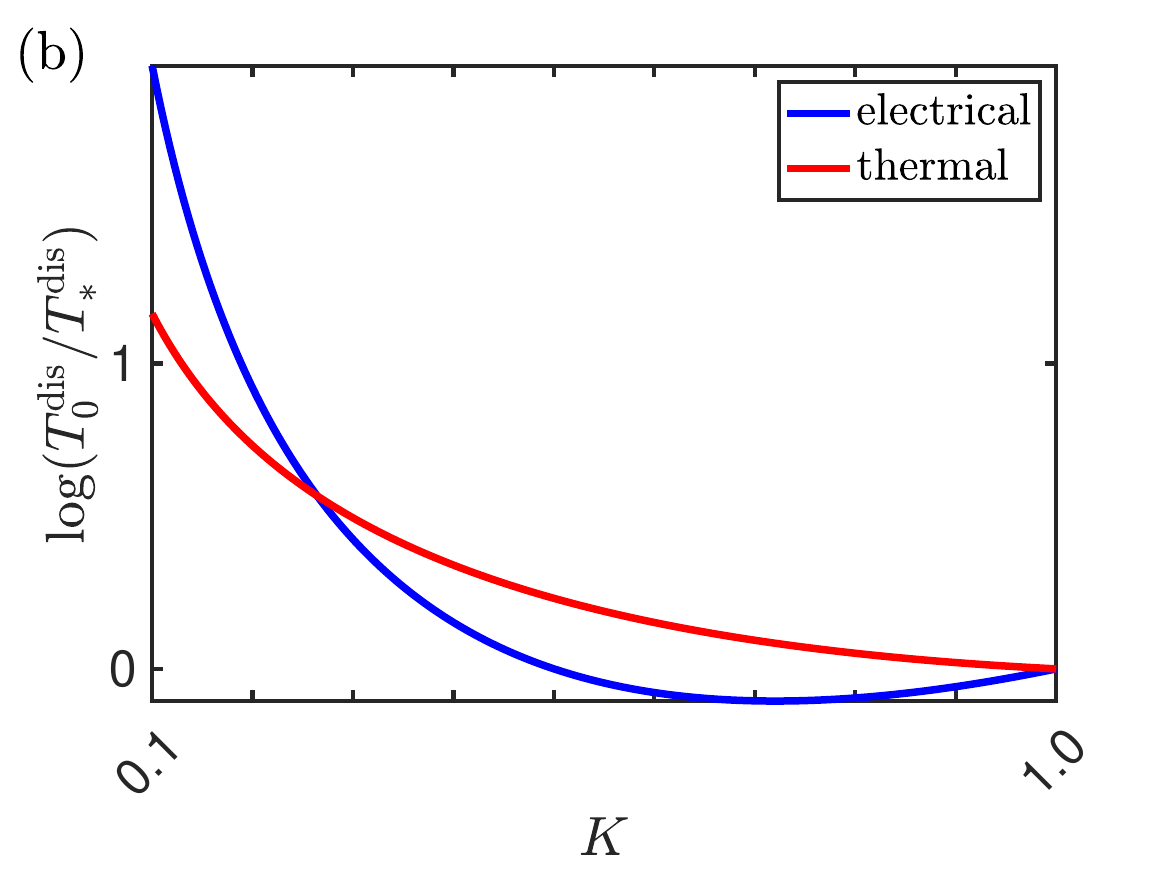}
\centering
\caption{ The crossover temperature of both electrical and thermal conductivites for (a) clean limit(umklapp scattering) and (b) dirty limit(disorder scattering).}
\label{betadis}
\end{figure}

Eqs.~\eqref{crossovertemperatureinteractions} and \eqref{crossovertemperatureinteractionsthermal} are the primary results of our paper.
These expressions show how the crossover temperature can vary with the strength of the repulsive interaction $K < 1$, relative to its value at the noninteracting fixed point.
We plot this dependence in Fig.~\ref{betadis}.
Observe that the variation can be more than an order of magnitude as the strength of the repulsive interaction is increased (i.e., as $K$ is decreased).
The span of this variation is greater for the umklapp scattering than the disorder scattering and greater for electrical conductivity relative to the thermal conductivity.
Because all the conductivities scale with a common factor of $T^{-2 K}$, which cancels out when obtaining \eqref{crossovertemperatureinteractions} and \eqref{crossovertemperatureinteractionsthermal}, the relative enhancement of the crossover temperature is entirely due to the interaction dependences of the {\it slopes} of the conductivities, rather than their exponents.

\section{Conclusion}
\label{conclusionsection}

We studied charge and heat transport in an interacting Luttinger liquid coupled to acoustic phonons using the memory matrix formalism. By explicitly incorporating the Debye cutoff in electron-phonon correlation functions, we computed the dc conductivities across low- and high-temperature regimes in both clean (umklapp-dominated) and dirty (disorder-dominated) limits.

We showed how strong correlations not only modify transport exponents---they also alter the crossover temperatures scale separating different scattering regimes.
Our central result, summarized in \eqref{crossovertemperatureinteractions} and \eqref{crossovertemperatureinteractionsthermal}, is that the crossover temperature $T_0$ between low- and high-temperature transport regimes is enhanced by repulsive electronic correlations: $T_0 \sim \Theta_D \times f(K)$, where $\Theta_D$ is the Debye temperature and the function $f(K)$ can increase (depending on the scattering type) by an order of magnitude as the strength of repulsive electron-electron interactions is increased.
This occurs as the Luttinger parameter $K$ is decreased from the free fermion point $K=1$ towards $K = 0$. 
The consequence is that the low-temperature scattering regime is extended to higher temperatures, with the would-be high-temperature regime potentially rendered unobservable.
It would be interesting to study whether some analog of this mechanism survives in models for two-dimensional strange metals.

While our one-dimensional model neglects spin degrees of freedom and vertex corrections, the mechanism we identify---interaction-enhanced crossover scales---may be general to strongly correlated systems.
It may be interesting to consider the crossover temperature in more general one-dimensional Luttinger liquids with more channels, which feature additional symmetries and robust low-temperature phases  \cite{Plamadeala2014,Plamadeala2016}. 
The extension to higher dimensions is of great interest: a particularly interesting and tractable model may be that of \cite{2024PhRvL.132w6501B}.

\acknowledgements
We thank Sankar Das Sarma, Hart Goldman, Steve Kivelson, Chao-Jung Lee, Sri Raghu, and J$\ddot{{\rm o}}$rg Schmalian for helpful discussions.
This work was supported by the U.S.~Department of Energy, Office of Science, Office of Basic Energy Sciences, under Award No.~DE-SC0026362. 
M.M.~acknowledges the kind hospitality of the Kavli Institute for Theoretical Physics, which is supported in part by the National Science Foundation under Grants No.~NSF PHY-1748958 and PHY-2309135.
\newpage
\appendix

\section{Static Susceptibilities}
\label{appendixsusceptibility}

The static susceptibility matrix $\hat{\chi}_{{\cal Q} {\cal Q}'}=G_{{\cal Q}{\cal Q}'}^{R}(\omega =0)/L$, where ${\cal Q}$ is a conserved current. 
We start with the electrical current,
\begin{align}
J_{e}&= {ev_{e}K}\int_{x}\Pi(x).
\end{align}
In general, the retarded Green's function $G_{{\cal Q}{\cal Q}'}^{R}(\omega)$ can be obtained by Wick rotating the corresponding Euclidean Green's function $G_{{\cal Q}{\cal Q}'}^{E}(i\omega_{E}\rightarrow \omega+i\delta)$, where $\omega_{E}$ is the Euclidean frequency. 
 The Euclidean Green's function is (setting $\delta\rightarrow 0$)
\begin{align}
\hat{\chi}_{J_{e}J_{e}} & \equiv {1 \over L} \lim_{\omega_{E} \rightarrow 0} \int_{\tau} e^{i\omega_{E}\tau}\langle J_{e}(\tau) J_{J}^{e}(0) \rangle = ({ev_{e}K})^{2} \lim_{\omega_{E} \rightarrow 0} {1 \over L} \int_{\tau,x,y} e^{i\omega_{E} \tau}\langle \Pi (x,\tau) \Pi(y,0) \rangle,
\end{align}
where the Euclidean time $\tau\in[0,1/T]$, and $\langle\cdots\rangle$ stands for thermal average. 
The Euclidean action $S_{E}$ is
\begin{align}
S_{E}=\int_{0}^{\beta}d\tau\int dx\left[-i\Pi\partial_{\tau}\phi+{\pi v_{e}K\over 2}\Pi^{2}+{2\over 2\pi K}(\partial_{x}\phi)^{2}\right].
\end{align}
The Fourier transform of $\phi(x,\tau)$ and $\Pi(x,\tau)$ are such that
\begin{align}
\phi(x,\tau)&={1\over\sqrt{\beta L}}\sum_{q,\omega_{n}}\phi(q,\omega_{n})e^{i(qx+\omega_{n}\tau)},\\
\Pi(x,\tau)&={1\over\sqrt{\beta L}}\sum_{q,\omega_{n}}\Pi(q,\omega_{n})e^{i(qx+\omega_{n}\tau)},
\end{align}
where $\omega_n=2\pi nT$ is the bosonic Matsubara frequency. The Euclidean action $S_{E}$ becomes
\begin{align}
S_{E}={1\over 2}\sum_{q,\omega_{n}}
\big(\phi(-q,-\omega_{n})~~\Pi(-q,-\omega_{n})\big)
{\cal G}^{-1}
\begin{pmatrix}
\phi(q,\omega_{n})\\
\Pi(q,\omega_{n})
\end{pmatrix},
\end{align}
with
\begin{align}
{\cal G}^{-1}(q,\omega_{n})=
\begin{pmatrix}
{v_{e}q^{2}\over \pi K}&-\omega_{n}\\
\omega_{n}&\pi v_{e}K
\end{pmatrix}.
\end{align}
Since ${\rm det}~{\cal G}^{-1}=\omega_{n}^{2}+v_{e}^{2}q^{2}$, the inverse matrix is such that
\begin{align}
{\cal G}(q,\omega_{n})={1\over \omega_{n}^{2}+v_{e}^{2}q^{2}}
\begin{pmatrix}
\pi v_{e}K&\omega_{n}\\
-\omega_{n}&{v_{e}q^{2}\over \pi K}
\end{pmatrix}.
\end{align}
The diagonal components give the propagators of $\phi$ and $\Pi$, respectively. Hence, we have
\begin{align}
\langle\Pi(q,\omega_{n})\Pi(-q,-\omega_{n})\rangle={1\over\pi v_{e}K}{v_{e}^{2}q^{2}\over\omega_{n}^{2}+v_{e}^{2}q^{2}}.
\end{align}
For the static susceptibility, take $\omega_{n}=0$, and $q\rightarrow 0$
\begin{align}
\chi_{\Pi\Pi}^{\rm static}=\lim_{q\rightarrow 0}\lim_{\omega_{n}\rightarrow 0}\langle\Pi(q,\omega_{n})\Pi(-q,-\omega_{n})\rangle={1\over \pi v_{e}K}.
\end{align}
As a result, we have
\begin{align}
\hat{\chi}_{J_{e}J_{e}}={e^{2}v_{e}K\over \pi}.
\end{align}
On the other hand, the thermal current $J_{T}$ and momentum deviation from the Fermi surface $J_{D}$ are given by 
\begin{align}
J_{T}&=-v_{e}^{2} \int_{x} \Pi(\partial_{x}\phi)-v_{p}^{2}\int_x P(\partial_{x}q),\\
J_{D}& = -{v_{e}^{2}} \int_{x} \Pi(\partial_{x}\phi)-{v_{e}^{2}}\int_x P(\partial_{x}q).
\end{align}
By defining the electron and phonon momentum operators,
\begin{align}
{\cal P}_{e}\equiv -\int_{x} \Pi(\partial_{x}\phi),\quad {\cal P}_{p}\equiv -\int_{x} P(\partial_{x}q).
\end{align}
So that the currents are becoming
\begin{align}
J_{T}& = v_{e}^{2}{\cal P}_{e}+v_{p}^{2}{\cal P}_{p},\\
J_{D}& = v_{e}^{2}{\cal P}_{e}+v_{e}^{2}{\cal P}_{p}.
\end{align}
Because the electron and phonon are decoupled in $H_0$, we have $\hat{\chi}_{{\cal P}_{e}{\cal P}_{p}}=0$, which makes
\begin{align}
\hat{\chi}_{J_{T}J_{T}}&=v_{e}^{4}\hat{\chi}_{{\cal P}_{e}{\cal P}_{e}}+v_{p}^{4}\hat{\chi}_{{\cal P}_{p}{\cal P}_{p}},\\
\hat{\chi}_{J_{D}J_{T}}=\hat{\chi}_{J_{T}J_{D}}&=v_{e}^{4}\hat{\chi}_{{\cal P}_{e}{\cal P}_{e}}+v_{e}^{2}v_{p}^{2}\hat{\chi}_{{\cal P}_{p}{\cal P}_{p}},\\
\hat{\chi}_{J_{D}J_{D}}&=v_{e}^{4}\hat{\chi}_{{\cal P}_{e}{\cal P}_{e}}+v_{e}^{4}\hat{\chi}_{{\cal P}_{p}{\cal P}_{p}}.
\end{align}
To obtain $\hat{\chi}_{{\cal P}_{c}{\cal P}_{c}}$, $c=e,p$. We start with a free massless boson with velocity $v$, the Hamiltonian can be diagonalized as
\begin{align}
H_{c}=\sum_{k}v_{c}|k|b_{k}^{\dag}b_{k},
\end{align}
and the total momentum ${\cal P}$ is
\begin{align}
{\cal P}=\sum_{k}kb_{k}^{\dag}b_{k}.
\end{align}
Since $[{\cal P},H]=0$, the static susceptibility is
\begin{align}
\hat{\chi}_{{\cal P}_{c}{\cal P}_{c}}={\beta\over L}\langle{\cal P}^{2}\rangle_{c}={\beta\over L}\sum_{k}k^{2}n_{B}(v_{c}|k|)[1+n_{B}(v_{c}|k|)],
\end{align}
where $n_{B}(x)=(e^{\beta x}-1)^{-1}$ is the phonon occupation. The thermodynamic limit gives 
\begin{align}
{\beta\over L}\sum_{k}k^{2}n_{B}(v_{c}|k|)[1+n_{B}(v_{c}|k|)]&=\beta\int\limits_{-\infty}^{\infty}{dk\over 2\pi}k^{2}{e^{\beta v_{c}|k|}\over (e^{\beta v_{c}|k|}-1)^{2}}={T^{2}\over \pi v_{c}^{3}}\int\limits_{0}^{\infty}dx{x^{2}e^{x}\over (e^{x}-1)^{2}}.
\end{align}
Using the famous result
\begin{align}
\int\limits_{0}^{\infty}dx{x^{2}e^{x}\over (e^{x}-1)^{2}}={\pi^{2}\over3},
\end{align}
we obtain
\begin{align}
\hat{\chi}_{{\cal P}_{c}{\cal P}_{c}}={\pi T^{2}\over 3v_{c}^{3}}.
\end{align}
Finally, 
\begin{align}
\hat{\chi}_{J_{T}J_{T}}&={\pi T^{2}\over 3}(v_{e}+v_{p}),\\
\hat{\chi}_{J_{D}J_{T}}=\hat{\chi}_{J_{T}J_{D}}&={\pi T^{2}\over 3}(v_{e}+{v_{e}^{2}\over v_{p}}),\\
\hat{\chi}_{J_{D}J_{D}}&={\pi T^{2}\over 3}(v_{e}+{v_{e}^{4}\over v_{p}^{3}}).
\end{align}

\section{Evaluation of $F^{i}_{m, {\cal Q}}$}
\label{commute}

In the limit $v_{p}\ll v_{e}$, $F^{U}_{m,{\cal Q}}=i[H^{U}_{m} , {\cal Q}]$, and $F^{\rm dis}_{m,{\cal Q}}={i\over\sqrt{D_{m}}}[H^{\rm dis}_{m} , {\cal Q}]$ are:
\begin{align}\label{correlator}
F_{m, J_{e}}^{U} & = A_{m, J_{e}}^{U} \int_{x} {1 \over a^{m}} \sin \big( \overline{k}_m x + m \phi \big) \big( \partial_{x} q \big),\\
F_{m,{\cal P}}^{U} & \approx A_{m, {\cal P}}^{U}\int_{x} {1 \over a^{m}} \cos \big( \overline{k}_m x + m \phi \big) \big( \partial_{x}^{2} q \big),\\
F_{m, J_{e}}^{{\rm dis}} & = {A_{m, J_{e}}^{{\rm dis}}\over \sqrt{D_{m}}} \left[ \int_{x} {1 \over a^{m}} \xi_{m}(x) e^{i m \phi } \big( \partial_{x} q \big) - h.c.\right], \\
F_{m,{\cal P}}^{{\rm dis}} & \approx {A_{m,{\cal P}}^{{\rm dis}} \over \sqrt{D_{m}}} \left[ \int_{x} {1 \over a^{m}} \xi_{m} e^{i m \phi} (\partial_{x}^{2} q)+ h.c. \right],
\end{align}
where ${\cal P}\in\{J_{T},J_{D}\}$.
The coefficients $A_{m,{\cal Q}}^{i}$ are 
\begin{align}
A_{m,J_{e}}^{U} &= i\pi ev_{e}K, \\
A_{m,J_{D}}^{U} &= v_{p}^{2}, \\
A_{m,J_{T}}^{U} &= v_{e}^{2}+v_{p}^{2},\\
A_{m, J_{e}}^{{\rm dis}} &= -ev_{e}K m,\\
A_{m, J_{D}}^{{\rm dis}} &= v_{e}^{2}, \\
A_{m, J_{T}}^{{\rm dis}} &= v_{p}^{2}.
\end{align}
The approximation indicated by $``\approx"$ is leading for $v_p \ll v_e$.

\section{Disorder Scattering Memory Matrix Details}
\label{appendixdisordermemory}

In this appendix, we give details for the calculation of the memory matrix in the disorder scattering limit presented in \S \ref{disordersection}.

The retarded Green's function ${1 \over L} \langle F_{m, J_{e}}^{\rm dis} , F_{m, J_{e}}^{\rm dis} \rangle_{\omega}$ is
\begin{align}
{1 \over L} \langle F_{m, J_{e}}^{\rm dis} , F_{m, J_{e}}^{\rm dis} \rangle_{\omega} & = { A_{m, J_{e}}^{\rm dis} A_{m, J_{e}}^{\rm dis} \over L D_{m}} \int\limits_{x,y,t} e^{i\omega t} g_{e}(x-y,t)g_{p}(x-y,t)\left[ \xi_{m}(x)\xi_{m}^{*}(y) + \xi_{m}^{*}(x)\xi_{m}(y) \right]. 
\end{align}
We perform an integration by parts and drop all boundary terms to obtain ${1\over L}\langle F_{m,{\cal P}}^{\rm dis};F_{m,{\cal P}'}^{\rm dis}\rangle_{\omega}$ for ${\cal P}, {\cal P}'\in\{J_{T},J_{D}\}$:
\begin{align}
{1 \over L} \langle F_{m,{\cal P}}^{\rm dis} , F_{m,{\cal P}'}^{\rm dis} \rangle_{\omega} & = -{ A_{m,{\cal P}}^{\rm dis} A_{m,{\cal P}'}^{\rm dis} \over LD_{m}} \int\limits_{x,y,t} e^{i\omega t} g_{e}(x,t)\big(\partial_{x}^{2}g_{p}(x,t)\big) \left[ \xi_{m}(x)\xi_{m}^{*}(y) + \xi_{m}^{*}(x)\xi_{m}(y) \right]. 
\end{align}
After disorder averaging, $\overline{\xi_{m}(x)\xi_{m}^{*}(y)}=D_{m}\delta(x-y)$, we obtain 
\begin{align}
{1 \over L} \langle F_{m, J_{e}}^{\rm dis} , F_{m, J_{e}}^{\rm dis} \rangle_{\omega} &=2A_{m,J_{e}}^{\rm dis} A_{m,J_{e}}^{\rm dis} G_{2}(\omega),\\
{1 \over L} \langle F_{m, {\cal P}}^{\rm dis} , F_{m, {\cal P}'}^{\rm dis} \rangle_{\omega} & = { 2A_{m, {\cal P}}^{\rm dis} A_{m, {\cal P}'}^{\rm dis} } \widetilde{G}_{2}(\omega).
\end{align}
The memory matrix $({\cal \hat{M}}^{\rm dis}_{m})^{{\cal P} {\cal P}'}$ is
\begin{align}
({\cal \hat{M}}^{\rm dis}_{m})^{J_{e}J_{e}}&=2A_{m, J_{e}}^{\rm dis}A_{m, J_{e}}^{\rm dis}{G_{2}(\omega)-G_{2}(\omega=0)\over i\omega}=\left.2A_{m,J_{e}}^{\rm dis}A_{m,J_{e}}^{\rm dis}\big(-i\partial_{\omega}G_{2}(\omega)\big)\right|_{\omega=0},\\
({\cal \hat{M}}^{\rm dis}_{m})^{{\cal P} {\cal P}'}&=2A_{m,{\cal P}}^{\rm dis}A_{m, {\cal P}'}^{\rm dis}{\widetilde{G}_{2}(\omega)-\widetilde{G}_{2}(\omega=0)\over i\omega}=\left.2A_{m, {\cal P}}^{\rm dis}A_{m, {\cal P}'}^{\rm dis}\big(-i\partial_{\omega}\widetilde{G}_{2}(\omega)\big)\right|_{\omega=0},
\end{align}
where the last equality is only valid when we take the dc limit.
$G_{2}(\omega)$ and $\widetilde{G}_{2}(\omega)$ are
\begin{align}
G_{2}(\omega)&=\int\limits_{x,t}e^{i\omega t}g_{e}(x,t)g_{p}(x,t)\delta(x),\\
\widetilde{G}_{2}(\omega)&=-\int\limits_{x,t}e^{i\omega t}g_{e}(x,t)\big(\partial_{x}^{2}g_{p}(x,t)\big) \delta(x).
\end{align}

We calculate:
\begin{align}
\left.-i\partial_{\omega}G_{2}(\omega)\right|_{\omega=0}&=\int\limits_{x,t}t\cdot g_{e}(x,t)g_{p}(x,t)\delta(x),\\
&={a^{2}\over 2v_{e}^{2}v_{p}}\big({2\pi Ta\over v_{e}}\big)^{2\Delta_m-2}e^{i(K-1/2)\pi}\big({2T\over v_{p}}\big)^{2}\int\limits_{0}^{\Theta_{D}/2T}dx {|\Gamma(\Delta_m+i{x\over\pi})|^{2}\over \Gamma(2\Delta_m)}{x\over \sinh(x)}, \cr
\left.-i\partial_{\omega}\widetilde{G}_{2}(\omega)\right|_{\omega=0}&=-\int\limits_{x,t}t\cdot g_{e}(x,t)\big(\partial_{x}^{2}g_{p}(x,t)\big) \delta(x)\\
&={a^{2}\over 2v_{e}^{2}v_{p}}\big({2\pi Ta\over v_{e}}\big)^{2\Delta_m-2}e^{i(K-1/2)\pi}\big({2T\over v_{p}}\big)^{4}\int\limits_{0}^{\Theta_{D}/2T}dx {|\Gamma(\Delta_m+i{x\over\pi})|^{2}\over \Gamma(2\Delta_m)}{x^{3}\over \sinh(x)}.\nn
\end{align}
The integral over $x$ is evaluated in the low ($T \ll \Theta_D$) and high ($T \gg \Theta_D$) temperature limits. 
At low temperatures,
\begin{align}
\int\limits_{0}^{\Theta_{D}/2T}dx {|\Gamma(\Delta_m+i{x\over\pi})|^{2}\over \Gamma(2\Delta_m)}{x^{\alpha}\over \sinh(x)}\approx {4\pi^{2-2\Delta_{m}}\over \Gamma(2\Delta_m)}\int\limits_{0}^{\infty}dx x^{2\Delta_{m}+\alpha-1}e^{-2x}={\pi^{2-2\Delta_m}\over \Gamma(2\Delta_m)}{\Gamma(2\Delta_m+\alpha)\over 2^{2\Delta_m+\alpha-2}}.
\end{align}
At high temperatures,
\begin{align}
\int\limits_{0}^{\Theta_{D}/2T}dx {|\Gamma(\Delta_m+i{x\over\pi})|^{2}\over \Gamma(2\Delta_m)}{x^{\alpha}\over \sinh(x)}\approx {|\Gamma(\Delta_m)|^{2}\over \Gamma(2\Delta_m)}\int\limits_{0}^{\Theta_{D}/2T}dx x^{\alpha-1}={|\Gamma(\Delta_m)|^{2}\over \Gamma(2\Delta_m)}{1\over\alpha}\left({\Theta_{D}\over 2T}\right)^{\alpha}.
\end{align}

Plugging these results in to the expression for the conductivity \eqref{transport} and using the static susceptibilities given at the start of \S \ref{umklappsection}, the electrical $\sigma$ and thermal $\kappa$ conductivities are found to be 
\begin{align}
\sigma\propto
\begin{cases}
T^{-2\Delta_{m}},&T \ll \Theta_D,\\
T^{1-2\Delta_{m}},& T \gg \Theta_D,
\end{cases}
\end{align}
\begin{align}
\kappa\propto
\begin{cases}
T^{1-2\Delta_{m}},& T \ll \Theta_D\\
T^{4-2\Delta_{m}},& T \gg \Theta_D .
\end{cases}
\end{align}

\bibliography{bigbib}

\end{document}